\documentclass[aps,pra,superscriptaddress,twocolumn,a4paper,nofootinbib]{revtex4-1}
	
\usepackage[utf8]{inputenc} \usepackage{amsmath} \usepackage{graphicx}
\usepackage[left=2.5cm,right=2.5cm,top=2cm,bottom=2cm]{geometry}
\usepackage{hyperref} \hypersetup{ pdfauthor = {Rojiar Penjweini,
    Markus Sondermann}, pdftitle = {Nonlinear optics with full 3D
    illumination}, }

\begin{document}

\title{Nonlinear optics with full three-dimensional illumination}

\author{Rojiar Penjweini} \email[]{rojiar.penjweini@mpl.mpg.de}
\affiliation{Friedrich-Alexander-Universit\"at Erlangen-N\"urnberg
  (FAU), Department of Physics, Staudtstr. 7/B2, 91058 Erlangen,
  Germany} \affiliation{Max Planck Institute for the Science of Light,
  Staudtstr. 2, 91058 Erlangen, Germany}

\author{Markus Weber} \affiliation{Friedrich-Alexander-Universit\"at
  Erlangen-N\"urnberg (FAU), Department of Physics, Staudtstr. 7/B2,
  91058 Erlangen, Germany} \affiliation{Max Planck Institute for the
  Science of Light, Staudtstr. 2, 91058 Erlangen, Germany}

\author{Markus Sondermann} \email[]{markus.sondermann@fau.de}
\affiliation{Friedrich-Alexander-Universit\"at Erlangen-N\"urnberg
  (FAU), Department of Physics, Staudtstr. 7/B2, 91058 Erlangen,
  Germany} \affiliation{Max Planck Institute for the Science of Light,
  Staudtstr. 2, 91058 Erlangen, Germany}

\author{Robert W. Boyd} \affiliation{Department of Physics, University
  of Ottawa, 25 Templeton Street, Ottawa, Ontario K1N 6N5, Canada}

\author{Gerd Leuchs} \affiliation{Friedrich-Alexander-Universit\"at
  Erlangen-N\"urnberg (FAU), Department of Physics, Staudtstr. 7/B2,
  91058 Erlangen, Germany} \affiliation{Max Planck Institute for the
  Science of Light, Staudtstr. 2, 91058 Erlangen, Germany}
\affiliation{Department of Physics, University of Ottawa, 25 Templeton
  Street, Ottawa, Ontario K1N 6N5, Canada}

\date{\today}

\begin{abstract}
We investigate the nonlinear optical process of third-harmonic
generation in the thus far unexplored regime of focusing the pump
light from a full solid angle, where the nonlinear process is
dominantly driven by a standing dipole-wave.  We elucidate the
influence of the focal volume and the pump intensity on the number of
frequency-tripled photons by varying the solid angle from which the
pump light is focused, finding good agreement between the experiments
and numerical calculations.  As a consequence of focusing the pump
light to volumes much smaller than a wavelength cubed the Gouy phase
does not limit the yield of frequency-converted photons.  This is in
stark contrast to the paraxial regime.  We believe that our findings
are generic to many other nonlinear optical processes when the pump
light is focused from a full solid angle.
\end{abstract}

\maketitle

\section{Introduction}
\begin{figure}[tb]
\centering \includegraphics[width=7.5cm]{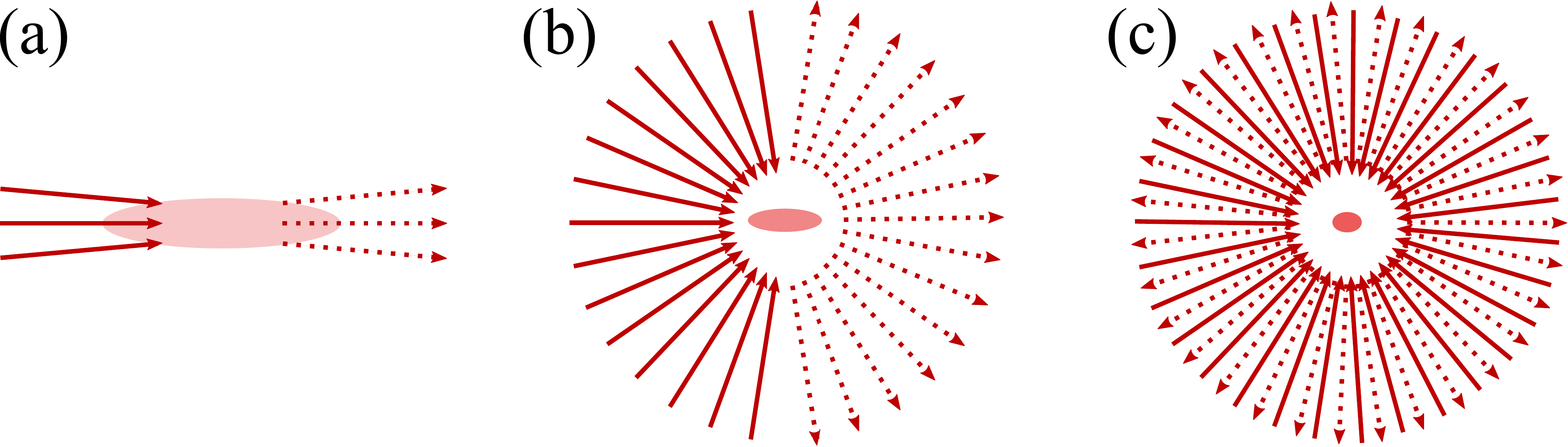}
\caption{\label{fig:1} Illustration of different focusing regimes: (a)
  paraxial regime using low NA. (b) non-paraxial regime of focusing
  with high NA. (c) focusing from a full solid angle.  Solid/dotted
  arrows represent the propagation direction of a wave coming
  toward/going out of the focus.  In non-paraxial regimes (b) and (c),
  the vector properties of the field are important.  The
  ros$\acute{\mathrm{e}}$ ellipse in the center indicates the size of
  the focal spot.  Note that in (c) the spot is not spherical which is
  a result of the vector properties of the light, not shown in the
  diagrams.  }
\end{figure}

The first multi-photon process was described by Göppert-Mayer when
calculating the spontaneous decay of the $2s$ state of the hydrogen
atom~\cite{goeppert-mayer1931}.  But it took until the invention of
the laser and the first investigation of second-harmonic generation by
\emph{Franken} and coworkers in 1961\,\cite{franken1961} before the
field of nonlinear optics took off.  Since then most of the
experiments have been performed in the paraxial regime with mildly
focused Gaussian beams, cf. Fig.~\ref{fig:1}a.  One can find only a
few reports on experiments where the pump light driving the nonlinear
process has been focused such that the paraxial approximation is not
valid~\cite{yew2003,biss2003,reichenbach2014,horneber2015,wang2016}.
These investigations treated second-harmonic generation at an
interface~\cite{yew2003,biss2003} or measured the nonlinear optical
response of nano
particles~\cite{reichenbach2014,horneber2015,wang2016}.  In a wider
sense, also multi-photon-excitation microscopy~\cite{zipfel2003} and
stimulated-emission depletion microscopy (STED)~\cite{westphal2005}
can be considered as nonlinear optics under non-paraxial conditions
when using microscope objectives with large enough numerical aperture
(NA) as depicted in Fig.~\ref{fig:1}b.  There is even one report on
STED in a 4Pi-microscopy setup using two microscope
objectives~\cite{dyba2002focal}, which however is still far away from
full $4\pi$ solid angle focusing.  Investigations in \emph{isotropic}
nonlinear media under clearly non-paraxial conditions are lacking.

Here, we investigate the nonlinear response under close to
full-solid-angle focusing, when the transmitted beam interferes with
the incoming beam to form a $4\pi$ standing wave
(cf. Fig.~\ref{fig:1}c).  Such a wave is a superposition of a
converging (inward moving) and a diverging (outward moving) dipole
wave~\cite{basset1986,cohen-tannoudji1989}.  In this standing wave
with a spherical phase front the wave vector of the pump naturally
averages to zero, while the broad spread of wave vectors has
implications on the phase matching of the nonlinear optical process,
as we will discuss later.  This is unlike the standing wave in a
cavity, where the wave vectors are typically confined within a small
cone.  Furthermore, the focal volume within a standing dipole-wave is
much smaller than a wavelength cubed~\cite{bokor2008,gonoskov2012}.
This is a new experimental regime in which no nonlinear optics
experiments have been conducted so far.  We perform such an experiment
in studying a paradigmatic process, third-harmonic generation (THG) in
an isotropic homogeneous medium (argon gas).

To focus the pump light from a full solid angle we use a parabolic
mirror (PM) with a focal length much shorter than its depth.  The PM
is illuminated with a mode that after reflection off the PM surface
resembles the radiation pattern of a linear electric
dipole~\cite{lindlein2007}.

Under such conditions, several questions as posed below arise.  In
order to address these questions, we briefly recall some essential
features of THG with Gaussian beams in the paraxial regime (see
e.g. Ref.~\cite{boyd2008nonlinear}): When the nonlinear medium is
longer than twice the Rayleigh length of the Gaussian beam and the
beam waist is located in the middle of the medium, the nonlinear
polarization induced in the interaction region before the beam waist
is 180 degree out of phase with the one generated in the diverging
beam behind the beam waist due to the Gouy phase.  Under conditions of
nominal phase matching this results in destructive interference of the
light fields generated in these two distinct half spaces.  One can
only compensate for the Gouy phase when choosing a nonlinear medium
with a positive phase mismatch.  Hence, in normally dispersive media
such as noble gases driven far from resonance, where the phase
mismatch is always negative, THG by four-wave mixing (FWM)
\renewcommand{\thefootnote}{$\ast \ast$} \footnote{By THG through FWM,
  we mean the process\\ $\omega + \omega + \omega \longrightarrow 3
  \omega$.} is not expected to occur (see e.g. section 2.10.3 in
Ref.~\cite{boyd2008nonlinear}).  Thus, the generation of
frequency-tripled photons can occur only as the result of higher-order
processes, and accordingly the observed power dependence is not of
third order (see e.g. Refs.\,
\cite{ganeev2006,ganeev2000,marcus1999,malcuit1990,lhuillier1988,ganeev1986,vaicaitis2009,vaicaitis2013}).

When focusing from a full solid angle with a dipole-like radiation
pattern, the full width at half maximum of the spatial intensity
distribution in the focus is on the order of the wavelength of the
pump light or smaller.  Consequently, the field amplitude of the pump
light varies from practically zero to its maximum value within a
wavelength and strictly violates the slowly-varying-amplitude
approximation that is inherent to the paraxial approximation.  Phase
matching in the paraxial regime is equivalent to velocity matching
such that the phase relation between pump, signal and idler is
preserved.  In the regime with light propagating in all directions
this concept no longer makes sense.  It will even turn out that over
the relevant length scale of the focal pump field distribution the
phase of the pump varies so weakly that it practically can be
neglected.

Therefore, one can ask the general questions: ``What determines the
efficiency of the nonlinear coupling in this extreme case?'', ``Which
predictions of the paraxial approximation are still valid in the
regime of extreme focusing'' and ``If one observes THG under this
experimental condition, how will the third-harmonic signal scale with
pump power and with the solid angle used for focusing?''

In this paper, we give answers to these questions.  In
Sec.~\ref{sec:exp} we describe our experimental apparatus and present
the experimental results.  As we will show, one indeed observes the
generation of frequency-tripled photons when focusing from a large
fraction of the full solid angle.  Based on our observations, we
identify six-wave mixing (SWM) as the underlying process, somewhat
resembling other experiments in the paraxial
regime~\cite{ganeev2000,vaicaitis2009,vaicaitis2013}.  Guided by this
finding we compare our experimental observations to numerical
simulations.  In the last section, we discuss our results and draw
some further conclusions.

\section{Experiment}
\label{sec:exp}
\subsection{Experimental setup}

\begin{figure*}[tb]
\centering\includegraphics[width=1\textwidth]{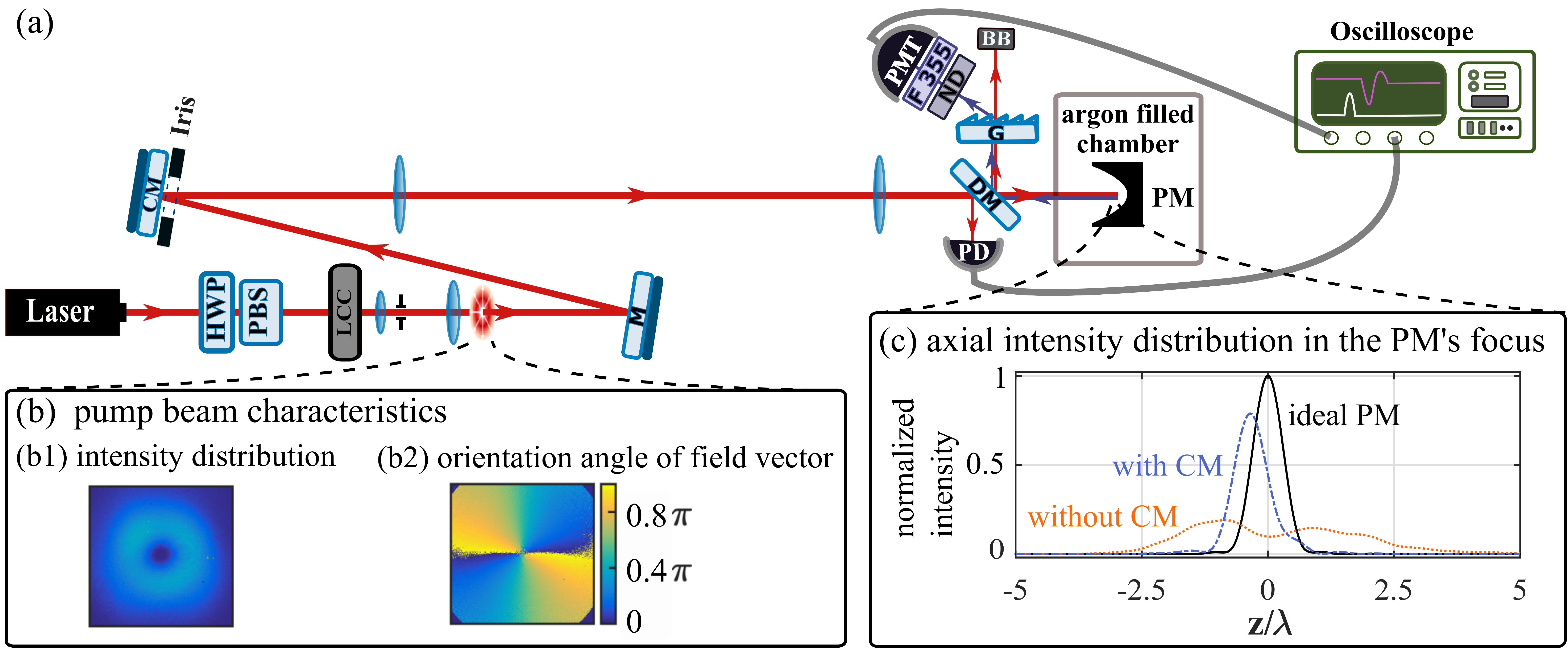}
\caption{\label{fig:2} (a) Scheme of the experimental setup. HWP:
  half-wave-plate, PBS: polarizing beam splitter, LCC: liquid-crystal
  polarization-converter, M: mirror, CM: compensation mirror, DM:
  dichroic mirror, PD: photodiode, PM: parabolic mirror, G: grating,
  BB: beam block, ND: neutral density filter, F 355: 355 nm laser-line
  filter, PMT: photomultiplier tube.  (b) Intensity distribution (b1)
  and spatially resolved orientation angle $\psi$ of the polarization
  vector (b2) of the pump beam.  $\psi = 0$ is pointing parallel to
  the optical table and perpendicular to the optical axis of the PM.
  (c) Simulated axial intensity distribution in the focal region of an
  aberration-free PM (black, solid curve), the PM used in the
  experiments (orange, dotted curve), and for this PM when the
  aberrations are partially corrected by use of the CM (blue,
  dash-dotted curve).  }
\end{figure*}

Figure~\ref{fig:2} shows a simplified scheme of our experimental
setup.  A pulsed Nd:YAG laser is used as the light source for the
fundamental beam.  This laser has a wavelength of 1064\,nm, a pulse
duration of 2\,ns, a repetition rate of 50\,Hz and a pulse energy of
up to 1\,mJ.  The beam power is adjusted by means of a half-wave-plate
and a polarizing beam-splitter.  The pump beam is transformed to a
radially polarized doughnut beam by passing the fundamental Gaussian
beam through a liquid-crystal polarization-converter (ARCoptix,
RADPOL).  Unwanted modes present in the beam leaving the
polarization-converter are rejected by means of a spatial filter.
Fig.~\ref{fig:2}b shows the intensity distribution and the
orientation-angle of the local polarization vector of the resulting
beam, both determined by a spatially resolved measurement of the
Stokes parameters~\cite{schaefer2007}.  This mode has an overlap with
the optimum mode for generating a linear dipole-wave in excess of
90\%.

The pump beam is aligned to the PM by using two mirrors such that the
beam propagates along the optical axis of the PM.  The PM is made of
aluminum and manufactured by single-point precision diamond-turning
(Kugler GmbH, Germany).  The focal length is $f=2.1\,\text{mm}$ and
the diameter of the exit PM's pupil is 20\,mm.  In addition, the PM
exhibits a bore hole of 1.5\,mm diameter at its vertex.  The PM is
placed inside a vacuum chamber which is first evacuated to the order
of $10^{-2}$\,mbar and then filled with argon gas.  The PM exhibits
deviations from a perfect parabolic shape, introducing significant
aberrations.  The aberrations are characterized by interferometric
measurements~\cite{leuchs2008}.  Based on the results of these
measurements, a compensation mirror (CM) was manufactured which
nominally imprints a wavefront modulation onto the incident beam that
is conjugate to the one imprinted by the aberrations of the PM.  The
actual wavefront imprinted by the CM is also determined by
interferometry.  The CM serves as one of the alignment mirrors
mentioned above (cf. Fig.~\ref{fig:2}a).  In order to avoid changes to
the imprinted wavefront occurring upon propagation, the electric field
distribution emerging from the CM is imaged 1:1 onto the entrance
aperture of the PM by means of a telescope.

We have assessed the impact of the PM's aberrations and the degree of
aberration compensation by the CM by simulating the focal intensity
distribution for various cases exploiting all available
interferometric data; see Fig.\ref{fig:2}c.  In comparison to an
aberration-free mirror, the PM alone exhibits a Strehl ratio of only
19\%, i.e., the maximum intensity in the focal region is about five
times below that observed for diffraction-limited focusing.  However,
the simulations predict that by using the CM the Strehl ratio can be
improved to 79\%.  This imperfect compensation is due to the fact that
the CM does not apply exactly the targeted phase distribution.

In the experiments presented below, we investigate the generation of
TH photons using different solid angles for focusing.  This variation
is achieved by aperturing the pump beam with an iris of adjustable
size.  The iris is positioned close to the CM and is thus likewise
imaged onto the PM aperture with the same telescope that is used for
imaging the CM.  By using the relation
$\tan{\vartheta/2}=r/2f$~\cite{lindlein2007} one can compute the
effective half-opening angle $\vartheta$ for a given iris radius $r$.
$\vartheta$ is then used for calculating the solid angle used for
focusing.

Because a strongly focused, radially polarized doughnut beam produces
an electric field distribution that closely resembles that of a linear
dipole oscillating along the optical axis of the focusing
device~\cite{quabis2000}, here we define the solid angle as the one
obtained when weighted with the angular intensity emission pattern of
a linear dipole: $\Omega=2\pi\int_{\vartheta_{min}}^{\vartheta_{max}}
\sin^{2}\vartheta \cdot \sin \vartheta d\vartheta$.  Using this
definition, $\Omega$ has an upper limit of $\frac{8\pi}{3}$ when
$\vartheta_{min}=0$ and $\vartheta_{max}=\pi$~\cite{sondermann2008}.
The specific geometry of our PM corresponds to $\vartheta_{min}$=
20$^{\circ}$ and $\vartheta_{max}\cong$134$^{\circ}$.  Therefore, the
maximum weighted solid angle covered by our PM is $0.94\times
\frac{8\pi}{3}$.

Frequency-tripled photons generated in the focal region are collimated
by the PM and afterwards reflected by a dichroic mirror (DM).  The
same DM directs a small fraction of the incident pump pulses onto a
photodiode (PD).  The PD signal serves as a trigger for detecting the
frequency-tripled beam.  To suppress any remaining pump light in the
detection path for the frequency-tripled photons, a grating (G)
separates this light from the frequency-tripled beam.  The pump beam
is finally dumped at a beam block (BB).  The frequency-tripled beam is
detected by a photomultiplier tube (PMT, Hamamatsu R11540) after
passing through a 355\,nm laser line filter and neutral density (ND)
filters.  The PMT cannot distinguish between pulses with different
photon numbers.  It rather responds nonlinearly to pulses with more
than one photon.  Therefore, the ND filters are chosen such that they
attenuate the frequency-tripled beam to an average photon number per
pulse smaller than unity.  In all experiments, the detected average
photon-number per pulse is obtained from a series of about 2500-3000
laser pulses focused by the PM.  Accounting for the attenuation
factors of the used ND filters and other optical losses (in total
$0.17\%$), we finally calculate the average number of
frequency-tripled photons generated in the focal region of the PM.

\subsection{Experimental results}
\label{subsec:experimental result}
At first, we check whether the generation of frequency-tripled photons
under strongly non-paraxial conditions is observed at all.  We monitor
the number of photons detected at a wavelength of 355\,nm while
varying the diameter of the iris limiting the beam size from 7\,mm to
20\,mm.  These diameters correspond to solid angles
$36\%\le\Omega/(8\pi/3)\le94\%$.  The smallest solid angle
investigated here corresponds to a half-opening angle of 80$^\circ$ or
a NA of 0.98, respectively.  Hence, all measurements are carried out
under strongly non-paraxial conditions.  For each iris diameter, the
laser power is adjusted such that the power transmitted through the
iris is the same.

\begin{figure}[tb]
\centering \includegraphics[width=7.5cm]{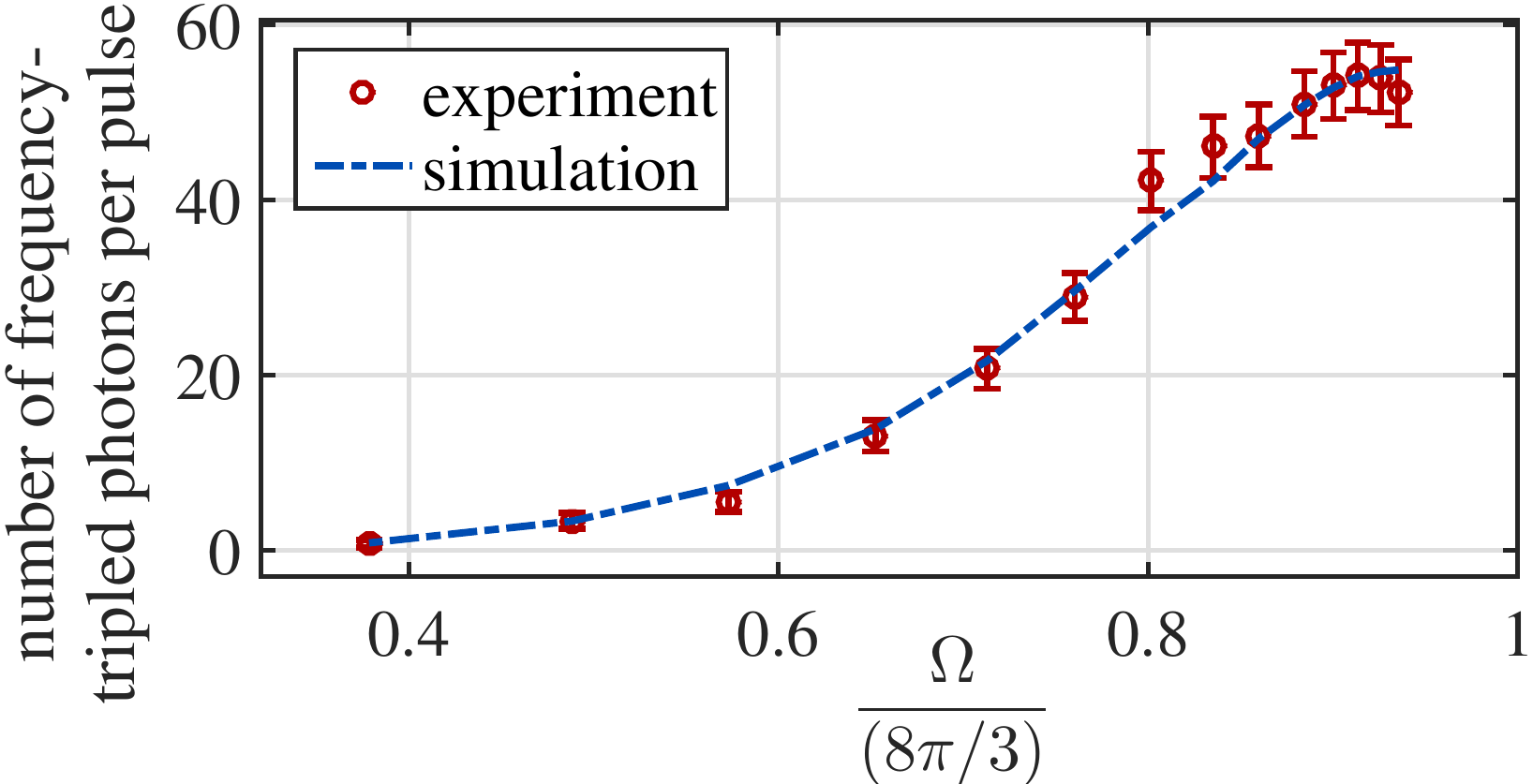}
\caption{\label{fig:3} Number of generated frequency-tripled photons
  versus the solid angle subtended by the pump beam.  Red points show
  the experimental results for a pressure of 668\,mbar and a fixed
  pump pulse energy of 114\,$\mu$J.  The dash-dotted blue line shows
  the result of a simulation when using the nonlinear susceptibility
  $\chi^{(5)}$ as a fit parameter (cf. Sec.~\ref{sec:sim} for
  simulation details).}
\end{figure}

\begin{figure}[tb]
\centering \includegraphics[width=7.5 cm]{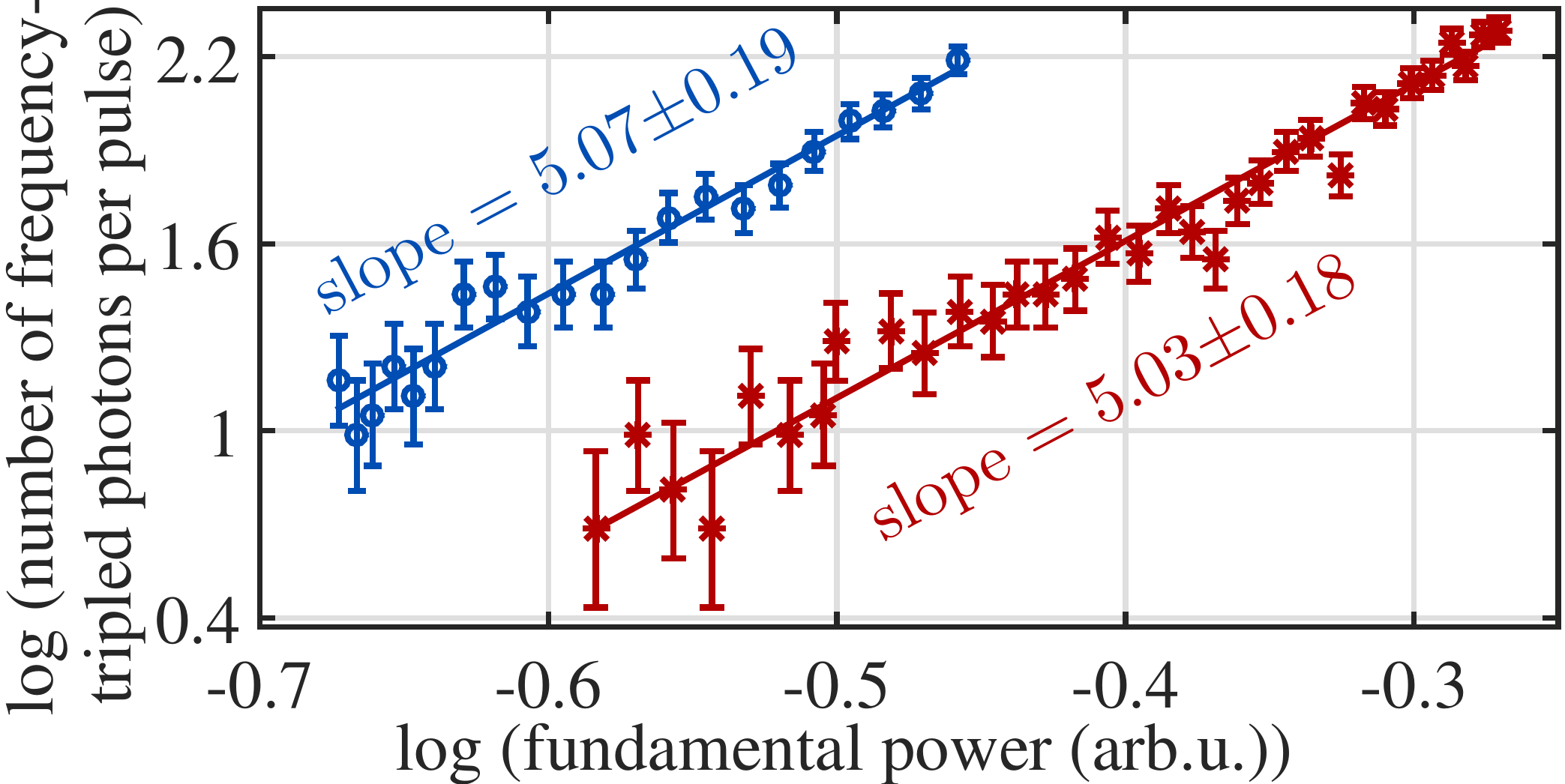}
\caption{\label{fig:4} Frequency-tripled photon generation vs. power
  of the fundamental beam at a pressure of 657\,mbar for two different
  solid angles: $55\%$ (red stars) and $94\%$ (blue circles) of full
  solid angle. The data is presented in double logarithmic scale
  (symbols).  The error bars are obtained from the Poisson statistics
  of the detected photons.  The line denotes the result of fitting a
  linear function to the data.}
\end{figure}

As is evident from Fig.~\ref{fig:3}, we indeed observe the generation
of frequency-tripled photons.  As one increases the solid angle used
for focusing, one observes a higher yield of frequency-tripled
photons.  The number of frequency-tripled photons per pulse is
affected by an interplay between the focal intensity and the focal
volume as the main nonlinear interaction region, as discussed below.
To investigate the generation of frequency-tripled photons in more
detail and to unveil the mechanism of the frequency conversion, we
measure the number of generated photons as a function of the peak
power of the fundamental beam.  Fig.~\ref{fig:4} shows the results for
two cases of focusing from $55\%$ and from $94\%$ of the full solid
angle.  A linear fit to the data in a double-logarithmic
representation produces a line with a slope of approximately 5 in both
cases.  Thus, the number of the generated frequency-tripled photons
scales with the fifth power of the optical power of the fundamental
beam.

This result answers one of the questions posed above: Under strongly
non-paraxial conditions, frequency-tripled photons are not generated
by a simple FWM process.  In the paraxial regime, dependences of the
frequency-tripled photon number on the pump power with orders ranging
from 3.5 to 5 have been reported
\cite{ganeev2006,ganeev2000,marcus1999,malcuit1990,lhuillier1988,ganeev1986,vaicaitis2009,vaicaitis2013}.
Possible explanations for this fifth-order dependence include a SWM
process~\cite{ganeev2000,vaicaitis2009,vaicaitis2013} and FWM with
phase matching achieved by by a Kerr
effect~\cite{ganeev1986,malcuit1990,ganeev2006}.  However, our
investigations exclude THG by FWM with phase matching enabled by the
Kerr effect.  As explained in detail in Appendix~\ref{sec:Kerr}, this
conclusion results from the fact that for argon gas and the pump beam
powers used in our experiment a positive phase mismatch cannot be
achieved via the Kerr effect.

Therefore, as done elsewhere for several experiments in the paraxial
regime~\cite{ganeev2000,vaicaitis2009,vaicaitis2013}, we attribute the
generation of frequency-tripled photons in our experiments to SWM: The
argon atoms absorb four photons at frequency $\omega$ and emit two
photons, one of them with frequency $\omega$ and the other one with
frequency $3\omega$.  Unlike for THG with focused light,
frequency-tripled generation through SWM is possible for both positive
and negative phase mismatch~\cite{kutzner1998}.  In either case, in
SWM one can compensate for the phase mismatch by suitable off-axis
wave vectors (see also Ref.~\cite{vaicaitis2009}).  When focusing from
large fractions of the solid angle, a broad spread of such wave
vectors is readily provided.

Having identified the dependence of the THG on pump power, we now
discuss the dependence of THG on solid angle in more detail: The
intensity in the focus of the PM is proportional to the solid angle
$\Omega$~\cite{sondermann2008}.  Therefore, for a SWM process as found
here one would expect the TH signal to scale with $\Omega^5$.
However, the experimental data underlying Fig.~\ref{fig:3} reveal a
slightly weaker dependence ($\Omega^4$).  This dependence results from
the fact that with increasing solid angle the focal volume decreases
(see simulations in Fig.~\ref{fig:7} of App.~\ref{sec:sim}).  The
decrease of focal volume is weaker than the increase of the peak
intensity in the focal region, resulting in the observed increase of
TH photons when focusing from a larger solid angle.

Building upon these arguments, we simulate frequency-tripled photon
generation, modeling the response of the medium with a fifth-order
susceptibility.  The detail of our theoretical simulation is explained
in App.~\ref{sec:sim}.

In Fig.~\ref{fig:3} we directly compare the simulation results to the
experimental data.  To within the experimental uncertainties we find a
good qualitative agreement between simulation and experiment.  By
fitting the simulation results to the experiment we obtain
$\chi^{(5)}=1.53^{+0.21}_{-0.18}\times10^{-48}\,(\text{m/V})^4/\text{bar}$
as the only fit parameter.  This value is of the same order of
magnitude as the value reported for the case of fifth harmonic
generation in Ref.~\cite{li1989} and therefore appears to be
reasonable.  The main uncertainty of the fitting procedure is given by
the accuracy of the pump power measurements, which is about 5\%.
Furthermore, in implementing our model numerically we made several
approximations, cf. App.~\ref{sec:prop}, which might influence the
uncertainty of the value obtained for $\chi^{(5)}$.  Nevertheless, we
conclude that our model yields a good agreement with the experimental
results.

\section{Discussion and conclusions}
\label{sec:discuss}
In the introduction of this paper we have raised several questions on
potential differences between harmonic generation in the paraxial
regime and when focusing the pump light from full solid angle.

Indeed, also in the latter case one observes the generation of
frequency-tripled photons in an isotropic medium with normal
dispersion.  As in the paraxial regime, one does not observe a
third-order dependence of the frequency-converted photons on pump
power as expected for a FWM process.  Rather, we have found a
fifth-order dependence which hints at SWM as the underlying process.

\begin{figure}
\centering \includegraphics[width=7.5cm]{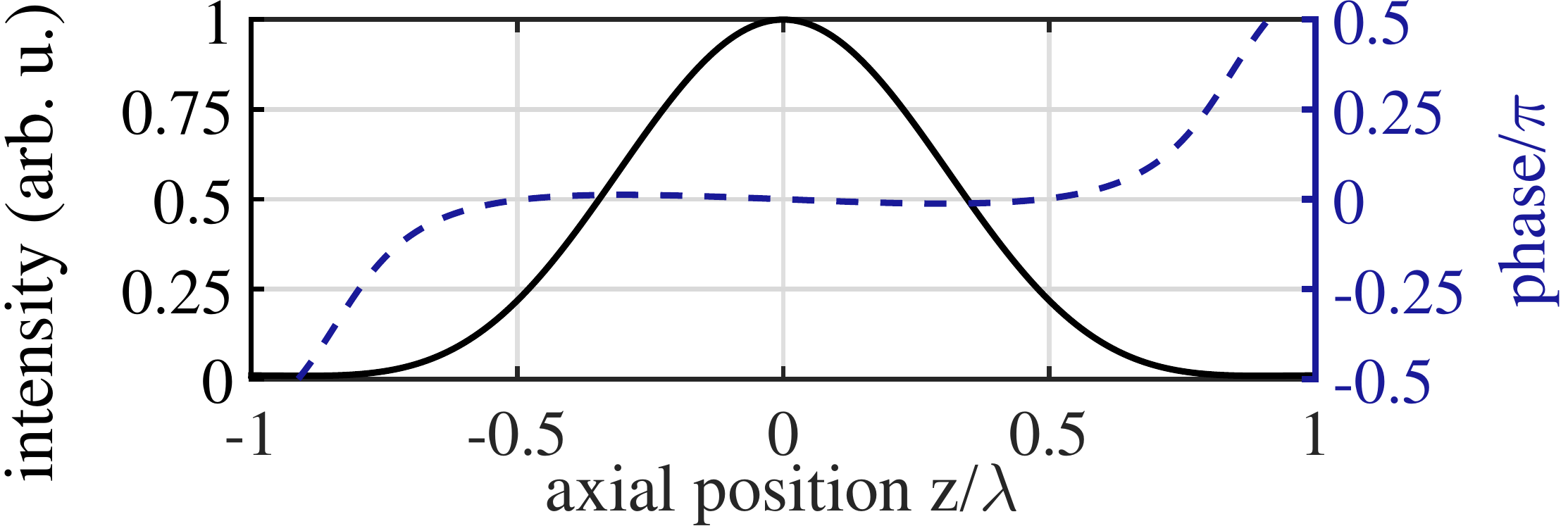}
\caption{\label{fig:5} Theoretical on-axis intensity (solid line) and
  phase distribution (dashed line) of the pump light on the optical
  axis of the PM when focusing with the complete PM covering $94\%$ of
  the full solid angle without any aberrations.}
\end{figure}

What is the origin of the suppression of the FWM contribution to THG
when focusing from a full solid angle?  The tempting answer might be
that the Gouy phase has the same detrimental effects as in the
paraxial regime.  And indeed, the standing spherical waves that are
generated by focusing from a full solid angle exhibit the Gouy phase,
i.e. a phase shift relative to a running spherical wave that emerges
from the focus~\cite{tyc2012}.  But being defined in such a way, the
Gouy phase does not reflect the total phase at a certain position in
the focal region.  As our simulations reveal, the spatial variation of
the phase of the pump field in the relevant focal region is not strong
enough to result in a complete suppression of a FWM signal.  As shown
in Fig.~\ref{fig:5}, the intensity distribution of the pump light
decays more quickly towards zero than the phase of the pump field
changes by $\pi/2$.  This is clearly different from what is found for
a focused Gaussian beam in the paraxial regime, for which a phase
distribution as the one displayed in Fig.~\ref{fig:5} can only be
found when choosing an unphysical beam waist.  The latter would
correspond to a lateral width-at-half-maximum that is smaller than the
minimal one obtainable in free space~\cite{gonoskov2012}.

Another possible reason for the suppression of FWM is found when
discussing wave vector diagrams
(e.g. Refs.~\cite{vaicaitis2009,boyd2008nonlinear}): For a FWM process
in a normally dispersive medium, there is no combination of three wave
vectors of the fundamental beam that results in a wave vector of the
TH light.  The wave vector mismatch is smallest for collinear wave
vectors.  In the experiment performed here, the strong focusing of the
fundamental beam induces a large spread of the directions of the
corresponding wave vectors.  This large spread results in larger wave
vector mismatches than in the paraxial regime.  Hence the FWM process
is suppressed even more strongly.  Contrarily, for the SWM process
focusing from full solid angle provides many possible combinations in
which five wave vectors of the fundamental beam can be matched to a
wave vector of the TH light.  We thus conclude that SWM is the
lowest-order process that can generate frequency-tripled photons in
the case of very tight focusing.

Although we investigated the influence of full-solid-angle focusing on
a specific nonlinear optical process, our findings -- especially the
ones about the role of the Gouy phase -- are valid for all nonlinear
optical processes, in particular to those of higher order.
Furthermore, collecting the generated photons over a full solid angle
minimizes losses and facilitates the investigation of the spatial
properties of many phenomena in nonlinear optics.

\section*{Acknowledgments}
The authors thank K.\,Mantel for the characterization of the
compensation mirror and M.\,Bader for discussions.  GL acknowledges
financial support from the European Research Council (ERC) via the
Advanced Grant 'PACART'.  RWB acknowledges support through the Natural
Sciences and Engineering Research Council of Canada and the Canada
Research Chairs program.


%

\onecolumngrid

\cleardoublepage

\appendix \twocolumngrid
\renewcommand\thefigure{\thesection.\arabic{figure}}

\section{Influence of four-wave mixing and Kerr effect}
\label{sec:Kerr}
\setcounter{figure}{0}

It is well known that THG by FWM with light focused such that the beam
waist lies in the middle of the interaction region is possible only if
the phase mismatch, $\Delta k=3k_{1}-k_{3}$, is
positive~\cite{boyd2008nonlinear}.  Here, $k_{1}$ is the wave number
of the fundamental beam and $k_{3}$ is the wave number of the TH beam.
In normal dispersive media (such as argon driven by 1064\,nm light),
with increasing frequency the refractive index increases and $\Delta
k=\frac{6\pi}{\lambda_{1}}(n^{(0)}_{1}-n^{(0)}_{3})$ is negative,
where $\lambda_{1}$ is the wavelength of the fundamental beam and
$n^{(0)}_{1}$ and $n^{(0)}_{3}$ are the linear refractive indices for
the fundamental beam and TH beam, respectively.  Thus, THG by FWM is
not possible in the normal dispersive media.

If THG is influenced by the Kerr effect, the phase mismatch will
become a function of intensity and nonlinear refractive indices as
\begin{equation}
\Delta k_{\text{Kerr}}=\frac{6\pi}{\lambda_{1}}
       [(n^{(0)}_{1}-n^{(0)}_{3})+(n^{(2)}_{1}-n^{(2)}_{3})I] \, ,
\label{eq:nonlineardeltak}
\end{equation}
with $n^{(2)}_{1}$ and $n^{(2)}_{3}$ denoting the \emph{nonlinear}
refractive indices for fundamental and TH beam.  $I$ is the intensity
of the fundamental beam.  The nonlinear refractive index for a single
intense fundamental beam with angular frequency $\omega$ is given
by~\cite{boyd2008nonlinear}
\begin{equation}
n^{(2)}_{1}=\frac{3}{4(n^{(0)}_{1})^{2}\epsilon _{0} c_{0}}
\chi^{(3)}\scalebox{0.85}{$(\omega=\omega+\omega-\omega)$} \, ,
\label{eq:n21}
\end{equation}
where $\epsilon _{0}$ is the vacuum permittivity, $c_{0}$ is the speed
of light in vacuum and $\chi^{(3)}$ is the third-order nonlinear
susceptibility.  The nonlinear refractive index for a weak
frequency-tripled beam with angular frequency $\omega'=3\omega$ in a
medium influenced by an intense fundamental beam with angular
frequency $\omega$ is~\cite{boyd2008nonlinear}
\begin{equation}
n^{(2)}_{3}=\frac{3}{2(n^{(0)}_{3})^{2}\epsilon _{0} c_{0}}
\chi^{(3)}\scalebox{0.85}{$(\omega'=\omega'+\omega-\omega)$} \, .
\label{eq:n23}
\end{equation}

Since the detuning of the pump as well as of the TH wave with respect
to the lowest excited state of argon has the same sign,
$\chi^{(3)}\scalebox{0.85}{$(\omega=\omega+\omega-\omega)$} $ and
$\chi^{(3)}\scalebox{0.85}{$(\omega'=\omega'+\omega-\omega)$}$ have
the same sign, too.  Obviously, then the same holds true for
$n^{(2)}_{1}$ and $n^{(2)}_{3}$.

From Eq.\,(\ref{eq:n21}) and Eq.\,(\ref{eq:n23}) we conclude that
\begin{flalign}
\begin{split}
& n^{(2)}_{1}-n^{(2)}_{3} = \\ & \frac{3}{4 \epsilon_{0} c_{0}} (
  \frac { \chi^{(3)}\scalebox{0.85}{$(\omega=\omega + \omega -
      \omega)$}}{(n^{(0)}_{1})^{2}}
  -\frac{2\chi^{(3)}\scalebox{0.85}{$(\omega'=\omega'+\omega-\omega)$}}{(n^{(0)}_{3})^{2}})
  .
\label{eq:dn2}
\end{split}
\end{flalign}
According to Eq.\,(\ref{eq:nonlineardeltak}), to get a positive
$\Delta k_{\text{Kerr}}$ and hence the possibility of THG for focused
light, two conditions should be fulfilled.  The first condition is
\begin{equation}
\label{eq:cond1}
n^{(2)}_{1}-n^{(2)}_{3} > 0
\end{equation}
and the second condition reads
\begin{equation}
\label{eq:cond2}
|(n^{(2)}_{1}-n^{(2)}_{3})I|>|(n^{(0)}_{1}-n^{(0)}_{3})| \, .
\end{equation}

We cannot check the first condition quantitatively, because, to the
best of our knowledge, the value of $\chi^
{(3)}\scalebox{0.85}{$(\omega'=\omega' + \omega - \omega)$}$ for a
strong beam at 1064\,nm and a weak beam at 355\,nm has not been
reported.  If $\chi ^ {(3)}\scalebox{0.85}{$(\omega=\omega + \omega -
  \omega)$}$ is sufficiently smaller than
$2\chi^{(3)}\scalebox{0.85}{$(\omega'=\omega' + \omega - \omega)$}$
such that the first condition is not fulfilled, then THG by FWM will
not be possible.  However, assuming that
$\chi^{(3)}\scalebox{0.85}{$(\omega=\omega+\omega-\omega)$}$ is
greater than $2\chi^{(3)}\scalebox{0.85}{$(\omega'=\omega' + \omega -
  \omega)$}$ and knowing that $n_{3}^{(0)}>n_{1}^{(0)}$, the first
condition given by Eq.~\ref{eq:cond1} is fulfilled.  With this
assumption, we check the second condition, setting
$n^{(2)}_{1}-n^{(2)}_{3} \cong n^{(2)}_{1}$ which is the case at which
the intensity $I$ needed to achieve the condition given by
Eq.~\ref{eq:cond2} is minimum.

Considering the linear refractive indices of argon at 1064\,nm and
355\,nm~\cite{rii}, we calculate
$n^{(0)}_{1}-n^{(0)}_{3}\cong-2.6\times 10^{-5}$.  For the fundamental
beam at 1064\,nm, the third-order nonlinear susceptibility of argon
gas is $\chi ^ {(3)}=7.8\times
10^{-27}\,(\text{m}^2/\text{V}^2)/\text{bar}$~\cite{li1989}.  The
maximum intensity which we reach just before the breakdown threshold
in argon gas is about $3.5\times 10^{13}\,\text{W}/\text{cm}^2$.  Our
experimental measurements are always done below the breakdown
threshold.  Setting $I$ to the intensity at the breakdown threshold
and using Eq.\,(\ref{eq:n21}) to calculate the nonlinear refractive
index, we conclude that $|(n^{(2)}_{1}-n^{(2)}_{3})I| \cong
|n^{(2)}_{1}I|=7.7\times 10^{-7}$ which is more than an order of
magnitude smaller than $|(n^{(0)}_{1}-n^{(0)}_{3})|$.  The difference
would be even more pronounced when $n^{(2)}_{1} \approx n^{(2)}_{3}$,
since then $|(n^{(2)}_{1}-n^{(2)}_{3})I|$ would be even smaller.
Therefore, even assuming most favorable conditions Eq.~\ref{eq:cond2}
cannot be fulfilled.  Thus we conclude that the generation of
frequency-tripled photons in our experiment is not the result of THG
by FWM, even when phase matching is influenced by the Kerr effect.

\section{Theoretical considerations}
\label{sec:sim}
\setcounter{figure}{0}

In what follows, we model the generation of frequency-tripled photons
in the focus of a parabolic mirror.

The electric field of the incident focused beam induces a nonlinear
polarization in the focal region of the PM.  The contribution of the
nonlinear polarization relevant for generating frequency-tripled
photons is $\textbf{P}_{3\omega}$.  In section
\ref{subsec:experimental result} we have argued that SWM is the
responsible process for the generation of photons with frequency
$3\omega$.  Hence, we write

\begin{align}
\begin{split}
\textbf{P}_{3\omega}&=\textbf{P}^{(5)}\\ &=5 \epsilon_{0} \chi^{(5)}
\scalebox{.85}{$(3 \omega=\omega+\omega+\omega+\omega-\omega)$}
\textbf{E}^4(\textbf{r})\textbf{E}^{*}(\textbf{r}) ,
\label{eq:p5}
\end{split}
\end{align}

where the factor 5 is the degeneracy factor, $\chi^{(5)}$ is the
fifth-order nonlinear susceptibility and $\textbf{E}(\textbf{r})$ is
the electric field of the focused fundamental beam.  $\textbf{r}=0$ is
the position of the geometrical focus of the PM.

Because $\textbf{P}_{3\omega}$ is a dipole-moment density, the dipole
moment oscillating at $3\omega$ that is induced in a volume element
$V_i$ is given by
\begin{equation}
\boldsymbol\mu_{3\omega,i}=\int_{V_i}\textbf{P}_{3\omega}
d^3\textbf{r}\, .
\label{eq:mu&p}
\end{equation}
In our simulations we associate $V_i$ with the volume of a unit cell
of the simulation grid.  The light emitted by each dipole is collected
by the parabolic mirror and propagates towards the detector.  The
detected signal is given by the interference of all these fields, with
the amplitude of the field emerging from $V_i$ being proportional to
$\mu_{3\omega,i}$.  We anticipate this interference process by
introducing an effective dipole moment
\begin{equation}
  \label{eq:M3omega}
  M_{3\omega}=\sum_i \gamma_i\cdot \mu_{3\omega,i}
\end{equation}
where the $\gamma_i$ are real weighting factors that account for the
projection onto a detection mode (see App.~\ref{sec:prop} for a
discussion on the influence of the spatial separation of the dipole
moments $\mu_{3\omega,i}$ onto the overall signal).

The total power that is radiated at frequency $3\omega$ by the dipole
moment $M_{3\omega}$ amounts to~\cite[Eq. 9.24]{jackson1999}
\begin{equation}
  W_{3\omega}=\frac{(3\omega)^4}{12 \pi \epsilon_{0}
    c_{0}^3}|M_{3\omega}|^2\, .
\label{eq:jackson}
\end{equation}
The medium is excited with pulses of Gaussian envelope of FWHM $\tau$.
Accounting for the observed 5$^\text{th}$-order dependence of the THG
photons on excitation power, the duration of the THG pulse is
$\tau/\sqrt{5}$.  Therefore, the number of frequency-tripled photons
per pulse becomes
\begin{equation}
N_{3\omega}=\frac{W_{3\omega}\tau}{3\sqrt{5}\hbar\omega}\, ,
\label{eq:N3}
\end{equation}
where $3\hbar\omega$ is the energy of a frequency-tripled photon.
Combining Eqs.~(\ref{eq:mu&p}) to (\ref{eq:N3}) we arrive at
\begin{equation}
  N_{3\omega}=\frac{225 \epsilon_0 \omega^3 \tau
    \chi^{(5)^{2}}}{4\sqrt{5}\pi\hbar c_{0}^3} \left| \sum_i
  \gamma_i\int_{V_i}\textbf{E}^4(\textbf{r}) \textbf{E}^*(\textbf{r})
  d^3\textbf{r}\right|^2 \, .
\label{eq:N3swm}
\end{equation}

The complex electric field $\textbf{E}(\textbf{r})$ in the focal
region of the PM is calculated by using the Debye integral
method~\cite{richards1959}.  By integrating over complex fields we
explicitly account for the spatial variation of the phase of
$\textbf{P}_{3\omega}$.  In our calculations we take into account the
measured aberrations of our PM, the measured phase-front induced by
the CM as well as the field distribution of the radially polarized
doughnut mode.  All aberrations are modeled as relative phases of the
electric field distribution incident onto the PM.  The pulse energy
and duration are the same as in the experiment underlying
Fig.~\ref{fig:3}.

For the nonlinear susceptibility $\chi^{(5)}$ there is -- to the best
of our knowledge -- no value reported in literature that was obtained
in a comparable experimental setting, i.e. the generation of the TH of
1064\,nm light by SWM.  Ref.~\cite{li1989} reports $\chi^{(5)}$ for
generating the fifth harmonic of 1064\,nm light, whereas some
$\chi^{(5)}$ values have been determined for SWM processes involving
(deep) ultraviolet light, see Ref.~\cite{kutzner1998} and references
therein.  Therefore, we here use $\chi^{(5)}$ as a fit parameter with
which we quantitatively match the outcome of the simulations to the
experimental results.

\begin{figure}[tb]
\centering\includegraphics[width=7cm]{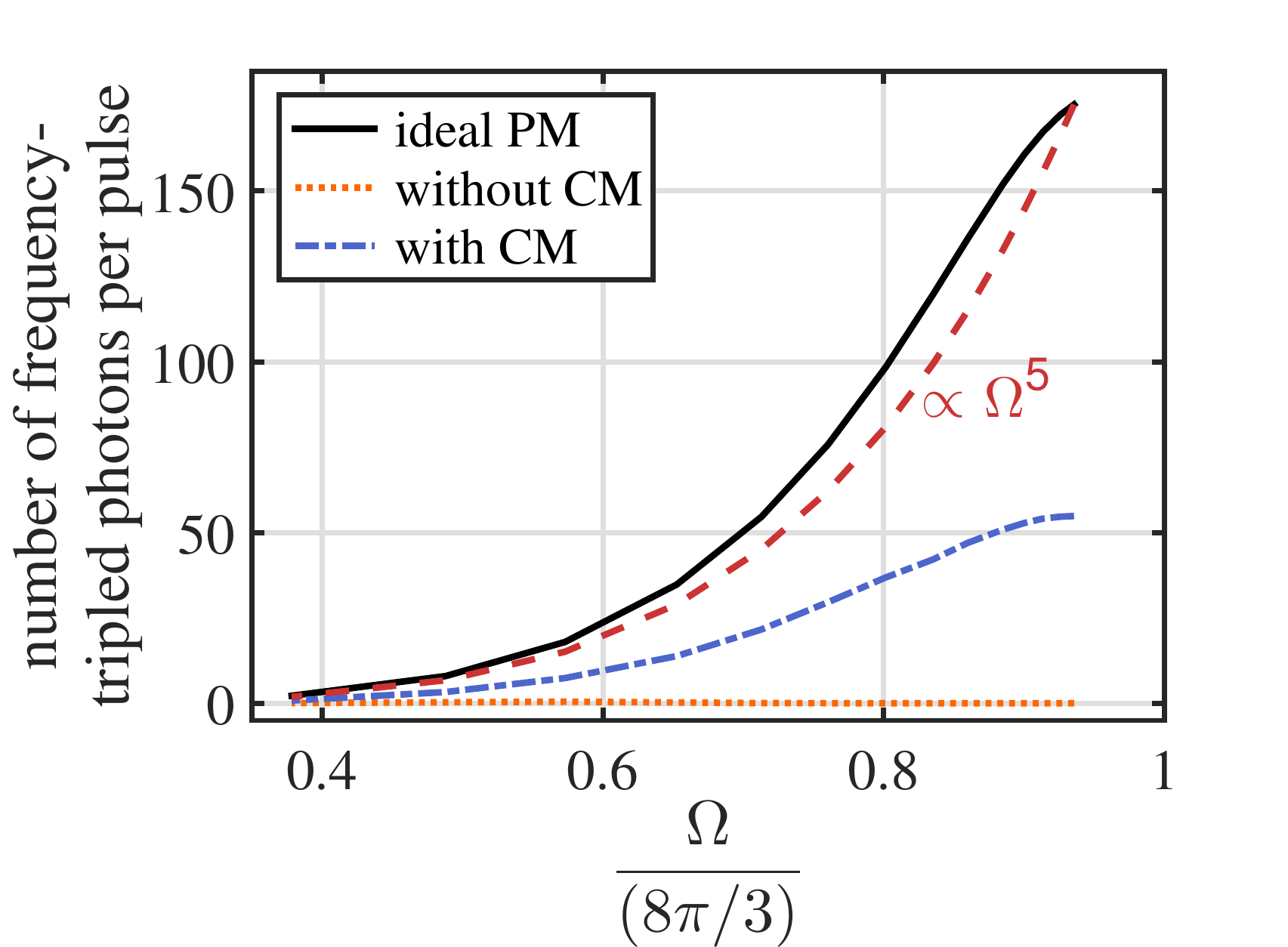}
\caption{\label{fig:6} Simulated frequency-tripled photon number vs
  solid angle used for focusing.  The dotted line is for the case of
  using our PM without correcting for its aberrations.  The
  dash-dotted line is for the case of employing a CM in our setup,
  which compensates the aberrations of our PM up to a Strehl ratio of
  79\%.  The solid line is for the case of an ideal PM without any
  aberrations.  The dashed line denotes a curve $\propto\Omega^5$ for
  comparison.  The absolute values of frequency tripled photons per
  pulse were obtained by fitting the case \lq with CM\rq\ to the
  experimental data.  }
\end{figure}

Fig.~\ref{fig:6} shows the result of simulating the generation of
frequency-tripled photons as a function of the solid angle used for
focusing.  We consider three cases, as shown in Fig~\ref{fig:6}.

\begin{figure*}[tb]
  \centering \includegraphics[width=14 cm]{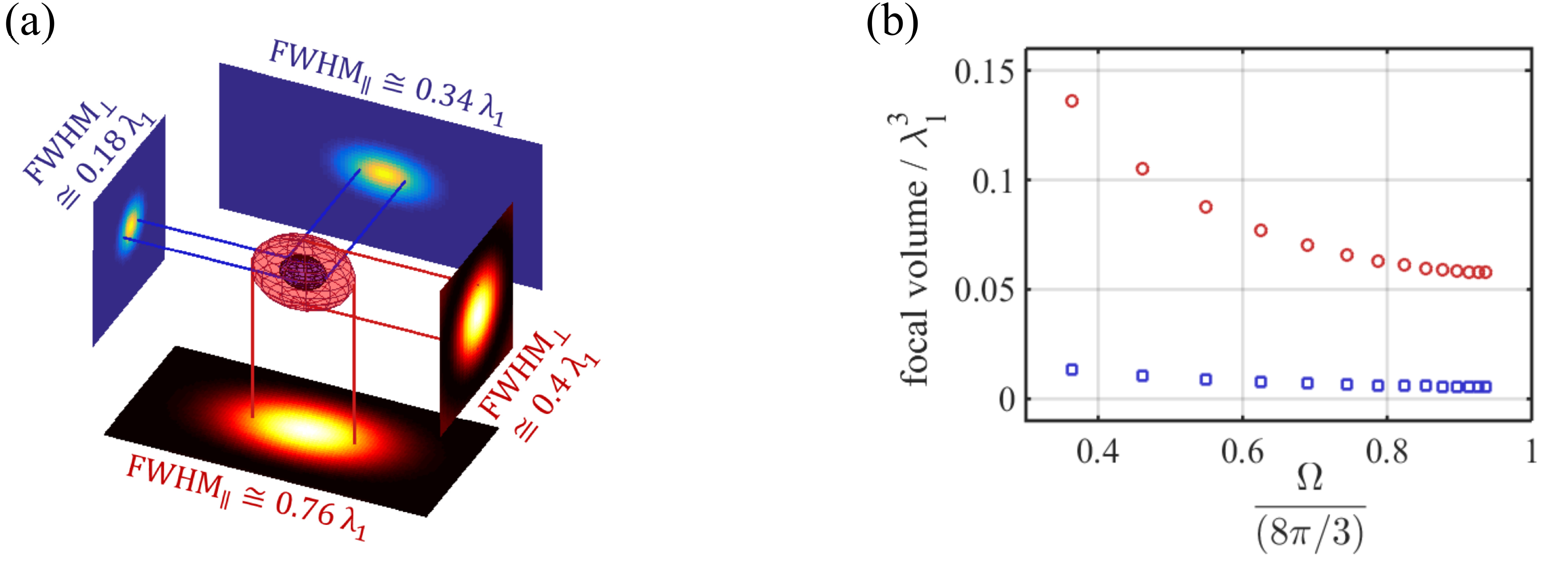}
\caption{\label{fig:7} (a) Simulated focal volume of the fundamental
  beam focused by the PM (outer spheroid) and the effective volume in
  which the frequency-tripled beam is generated through SWM (inner
  spheroid).  Aberrations of the PM are corrected for by using the CM.
  To obtain the respective volumes the FWHMs of the corresponding
  intensity distributions are determined.  The effective volume of
  frequency-tripled generation is determined from the fifth power of
  the intensity distribution of the fundamental beam.  The major axis
  of each spheroid equals the $\mathrm{FWHM}_{||}$ of the
  corresponding distribution along the axis of the PM and the minor
  axis is the $\mathrm{FWHM}_{\bot}$ of the distribution in the focal
  plane perpendicular to the axis of the PM.  (b) Focal volume
  vs. solid angle: The red circles correspond to the focal volume of
  the fundamental beam focused by PM. The blue squares correspond to
  the effective volume for frequency-tripled generation. All values
  are normalized to $\lambda_1^3$ where $\lambda_1$ is the wavelength
  of the fundamental beam.}
\end{figure*}

In the first case, we model a PM without any aberrations, i.e. at the
diffraction limit.  The number of photons at frequency $3\omega$ grows
monotonically with increasing solid angle.  This result can
intuitively be understood from the fact that the focal intensity of
the pump beam scales linearly with the solid angle
$\Omega$~\cite{sondermann2008}.  One would thus expect the conversion
efficiency of an $N$-th order process to scale with solid angle as
$\Omega^N$.  However, our simulation results do not show this
$\Omega^5$ dependence.  We attribute this different result to the
complicated spatial distribution of $\textbf{E}(\textbf{r})$ in the
focal region: for increasing solid angle, the maximum intensity of the
pump field in the focal region grows with $\Omega$.  However, the
focal volume shrinks when increasing the solid angle (see simulations
in Fig.~\ref{fig:7}).  Also the spatial distribution of the phase of
$\textbf{E}(\textbf{r})$ changes upon varying $\Omega$.  All these
effects result in the behavior observed in the simulation.

As a second case, we model the PM used in the experiments without any
aberration compensation, the number of generated frequency-tripled
photons is very low at each investigated solid angle,
cf. Fig.~\ref{fig:6}.  The maximum of $\sim0.5$ photons per pulse
occurs at a solid angle of $\Omega=0.57\cdot8\pi/3$.  That is, the
steady increase of the photon number for increasing $\Omega$ is no
longer observed.  The latter observation can be explained by the
spatial distribution of the aberrations over the surface of the PM.
Similar effects are also observed for other parabolic
mirrors~\cite{alber2017}.  Such aberrations appear to be typical for
deep parabolic mirrors, and seem to represent the current state of the
art.

Finally, as a third case we calculate the number of frequency-tripled
photons for the case of compensating the aberrations of the PM with a
CM, cf. Sec.~\ref{sec:exp}.  This case is used to fit the simulations
to the experimental results with $\chi^{(5)}$ as the only fit
parameter.  The simulation yields a steady increase of the photon
number with increasing solid angle.  Despite some saturation behavior
at solid angle fractions beyond 90\%, the results for the combination
PM+CM shows \emph{qualitative} similarities with the
diffraction-limited case.  However, the absolute photon numbers are
considerably smaller than in the diffraction limited case.  This
latter observation is readily explained by the still non-optimum
aberration compensation, which is expressed through a Strehl ratio of
79\%.  In the case of a nonlinear optical process as investigated
here, the influence of a non-unit Strehl ratio should exponentiate to
the order of the nonlinear process.  For the largest solid angle used
for focusing and for a fifth-order process, the simulation results
approximately exhibit this behavior.

\section{Collecting third-harmonic signals from spatially separated dipoles}
\label{sec:prop}
\setcounter{figure}{0}

In typical nonlinear optics experiments the light generated in a
wave-mixing process is collected from an extended spatial region.
This necessitates the account of the relative phases of the electric
fields generated at different positions when calculating the total
power that is generated in the nonlinear process.  There are two
contributions to the relative phases.  One stems from the relative
phase of the local pump field, which determines the phase of the
nonlinear polarization.  This contribution is directly included in our
simulations, cf. Eq.~\ref{eq:N3swm}.  The second contribution is
determined by the optical path-length difference (OPD) from the
different source dipoles in the nonlinear medium to the point of
detection.  We now discuss how to account for this contribution in our
particular scenario.

Whereas the OPD is readily defined in an experiment in which the
detection occurs only under a small solid angle, the situation is more
complicated when collecting light over the full solid angle.  For two
sources separated by a distance $d$ the OPD to a point of observation
lying on a circle with radius $\gg d$ is given by
\begin{equation}
  \label{eq:OPD}
  \text{OPD}=d\cdot\cos\vartheta
\end{equation}
with $\vartheta$ the angle to some reference direction.  Thus, there
is no unique OPD that is valid along all of the directions defined by
the wave vectors of the dipolar emission of two separated sources.
Moreover, the OPD is zero when averaging over $\vartheta$.

However, when collecting light over the entire solid angle with a deep
PM as in this work, the position of the light source determines the
phase front of the mode that is reflected off the parabolic surface.
These phase fronts can be expressed in terms of misalignment
functionals, which in general have to be calculated numerically by ray
tracing~\cite{leuchs2008}.  For the experimental scenario treated
here, we can make some simplifying assumptions that lead to analytic
expressions.

First, the focused pump field is predominantly polarized parallel to
the optical axis of the PM.  Thus the nonlinear polarization and
consequently all induced dipole moments oscillating at $3\omega$ are
oriented along this direction.  Since the extent of the focus is on
the order of a wavelength or even smaller, we assume that the
intensity distribution of the emission of all these dipoles is the
same as the one for a dipole located at the geometric focus of the
mirror.  After collimation by the PM and ignoring an overall amplitude
factor this intensity distribution reads~\cite{lindlein2007}
\begin{equation}
  \label{eq:intDipole}
I(r)\propto\frac{(r/f)^2}{[(r/f)^2/4+1]^4}
\end{equation}
with $f$ the PM focal length and $r$ the distance of a point in the
aperture plane of the PM to the optical axis.

Second, the simulations of the focal intensity distribution of the
pump light (cf. Sec.~\ref{sec:sim}) reveal that the electric field
$\mathbf{E}(\mathbf{r})$ is effectively concentrated in a narrow
region along the optical axis of the PM.  Since we observe that the
generation of frequency-tripled photons is proportional to the fifth
order of the pump power, we examine the distribution of
$|\mathbf{E}(\mathbf{r})|^{10}$.  We find that the half-width at
half-maximum of this distribution in lateral direction is about
$0.1\lambda_1$ for using the full mirror.  In the axial direction the
width is slightly larger.  For somewhat smaller solid angles, as was
the case in our measurements, the focal field distribution elongates
along the optical axis while the lateral extent is practically
constant.  We therefore infer that the phase fronts of the TH light
collected by the PM are mainly influenced by the axial position of the
emitters and that phase front distortions due to lateral displacements
can be neglected.  Identifying $\vartheta$ in Eq.~\ref{eq:OPD} with
the emission angle of the dipole radiation pattern, the optical
path-length difference of the emission from a dipole after collimation
by the PM can be written as
\begin{equation}
  \label{eq:front}
  \text{OPD}_i(r)=z_i ~ \frac{1-(r/2f)^2}{1+(r/2f)^2}\quad ,
\end{equation}
where $z_i$ is the axial displacement of the induced dipole
$\mu_{3\omega,i}$ from the PM focus.

For calculating the interference of the fields emitted by all dipoles
$\mu_{3\omega,i}$ in the focal region, we project each field
distribution on a detection mode.  We take the detection mode to be
the field distribution that is emitted by the largest dipole moment.
This dipole is located where the amplitude of the pump field is
maximum, the corresponding axial coordinate is $z_\text{max}$.  Then,
the overlap of the emission from dipole $\mu_{3\omega,i}$ with the
detection mode reads
\begin{equation}
  \label{eq:gammai}
  \gamma_i=\frac{\int I(r)\cdot\cos\left( \frac{6\pi}{\lambda_1}
    (z_i-z_\text{max})\frac{1-(r/2f)^2}{1+(r/2f)^2} \right)\,rdr}
        {\int I(r)\,rdr}\quad,
\end{equation}
with the integration performed over the entire aperture of the PM.
This is the factor $\gamma_i$ employed in the calculation of the
number of frequency-tripled photons in Eq.~\ref{eq:N3swm} in the main
part.

\end{document}